*Article*

# Laser-Inscribed Diamond Waveguide Resonantly Coupled to Diamond Microsphere

**Nurperi Yavuz** [1], **Mustafa Mert Bayer** [1,2] **Hüseyin Ozan Çirkinoğlu** [1], **Ali Serpengüzel** [1], **Thien Le Phu** [3,4], **Argyro Giakoumaki** [3,4], **Vibhav Bharadwaj** [3,4], **Roberta Ramponi** [3,4] and **Shane M. Eaton** [3,4,*]

[1] Department of Physics, Microphotonics Research Laboratory, Koç University, Rumelifeneri Yolu, Sarıyer, Istanbul 34450 Turkey; nyavuz16@ku.edu.tr (N.Y.); bayerm@uci.edu (M.M.B.); hcirkinoglu15@ku.edu.tr (H.O.Ç); aserpenguzel@ku.edu.tr (A.S.)

[2] Department of Electrical Engineering and Computer Science, University of California, Irvine, Irvine, CA 92697 USA

[3] IFN (Institute for Photonics and Nanotechnologies)—CNR, Piazza Leonardo da Vinci 32, Milano 20133 Italy

[4] Department of Physics, Politecnico di Milano, Piazza Leonardo da Vinci 32, Milano 20133 Italy; thien.lephu@polimi.it (T.L.P.); argyrogiak@gmail.com (A.G.); vibhavbharadwaj@gmail.com (V.B.); roberta.ramponi@polimi.it (R.R.)

* Correspondence: shane.eaton@gmail.com

**Abstract:** An all-diamond photonic circuit was implemented by integrating a diamond microsphere with a femtosecond-laser-written bulk diamond waveguide. The near surface waveguide was fabricated by exploiting the Type II fabrication method to achieve stress-induced waveguiding. Transverse electrically and transverse magnetically polarized light from a tunable laser operating in the near-infrared region was injected into the diamond waveguide, which when coupled to the diamond microsphere showed whispering-gallery modes with a spacing of 0.33 nm and high-quality factors of $10^5$. By carefully engineering these high-quality factor resonances, and further exploiting the properties of existing nitrogen-vacancy centers in diamond microspheres and diamond waveguides in such configurations, it should be possible to realize filtering, sensing and nonlinear optical applications in integrated diamond photonics.

## 1. Introduction

Silicon and silica are the most fundamental optical components in optical fibers, optical filters, optical amplifiers and photodetectors in integrated optical devices and photonic integrated circuits [1]. Such applications in the quantum regime have been rapidly developing, and there is a continuous search for a more suitable and faster photonics platform to replace silicon [2]. Having outstanding mechanical properties like hardness, high thermal conductivity and chemical inertness [3], as well as unique optical properties, such as a high Raman gain, large energy difference between valance and conduction bands (5.5 eV), and wide transmission window from ultraviolet to far-infrared (far-IR) region points out diamond as a promising photonic platform for optical filtering, sensing, amplification [4] and various quantum realm applications when integrated with quantum emitters [2].

The broad-band transparency of diamond with a refractive index of 2.4 [5] in the near-IR region facilitates the observation of whispering gallery modes (WGMs) between a closely spaced diamond waveguide (WG) and a diamond microsphere. WGMs manifest themselves by exploiting the total internal reflection due to the difference in refractive index between the external medium and the circular cavity, where the circumnavigating light realizes a localization around the microsphere [6]. As is well known, these localized modes exhibit high quality factor (Q-factor) resonances inside small spherical cavities. Previously, resonances from different circular microresonators such as silicon [7], silica [8] and diamond [4] were harvested for optical filtering [9], channel dropping [10], lasing [11] and sensing [12,13]. Non-resonant microspheres were also utilized for enzyme immobilization [14] and protein detection [15].

Moreover, to realize high Q-factor resonances, it is essential to achieve a superb coupling of light from guiding propagation media—such as tapered fibers [8], prisms [16], optical fiber-half-couplers (OFHCs) [17], and optical WGs [18]—to the smooth curvature of a microsphere. Among all, the femtosecond (fs)-laser photo-inscription technique enables the diamond to function as a WG for photonic integration to the diamond microsphere to achieve an all-diamond optical platform [19]. It was previously demonstrated that fs-laser-written diamond WGs have the capability of triggering high Q-factor WGMs inside silicon microcavities with high impact parameters, while exhibiting a polarization selective response for transverse electric (TE)- and transverse magnetically (TM)-polarized light [18]. Additionally, it is possible to excite the Fabry–Pérot (FP) resonances in the transmission direction in the diamond WG, which will further yield an integrated FP resonator along with a spherical resonator for the combined diamond photonic system.

In this work, we demonstrate an all-diamond photonic circuit, which performs the excitation of a 1 mm diamond microsphere via a fs-laser-written shallow diamond WG. The diamond WG has a depth of 20 μm, where the light propagates in a single mode (SM) with an elliptical mode shape. We observed high Q-factor WGMs, in the order of $10^5$, with a mode spacing of 0.33 nm, as well as FP resonances of the diamond WG, which have a free spectral range (FSR) of 87 pm, and Q-factors in the order of $10^4$. Spherical diamond microresonators on fs-laser-written diamond WGs show promise as novel all-diamond integrated photonic architectural components. By carefully engineering these high-quality factor resonances, and further exploiting the properties of existing nitrogen-vacancy centers in diamond microspheres and diamond waveguides in such configurations, it should be possible to realize filtering, sensing and nonlinear optical [20,21] applications in integrated diamond [22, 23] photonics.

## 2. Results and Discussions

For our applications, it is essential to have a WG mode close as possible to the surface of the diamond sample to enhance the evanescent coupling to the diamond optical resonator. However, because the laser fluence to form a Type II WG is greater than the surface ablation threshold, there is a limit to how close one can form a WG below the surface. The minimum depth we could achieve while maintaining good WG transmission was $d$ = 20 μm, measured from the surface to the center of the WG modification. The WG exhibited a 12.4 dB insertion loss at a 1550 nm wavelength, with a mode field diameter (MFD) of 16 μm × 20 μm, elongated in the vertical axis, which was measured with a similar WG characterization setup as in [24]. The cross-sectional microscope image and near-field mode profile of the optimum WG are shown in Figure 1 [25].

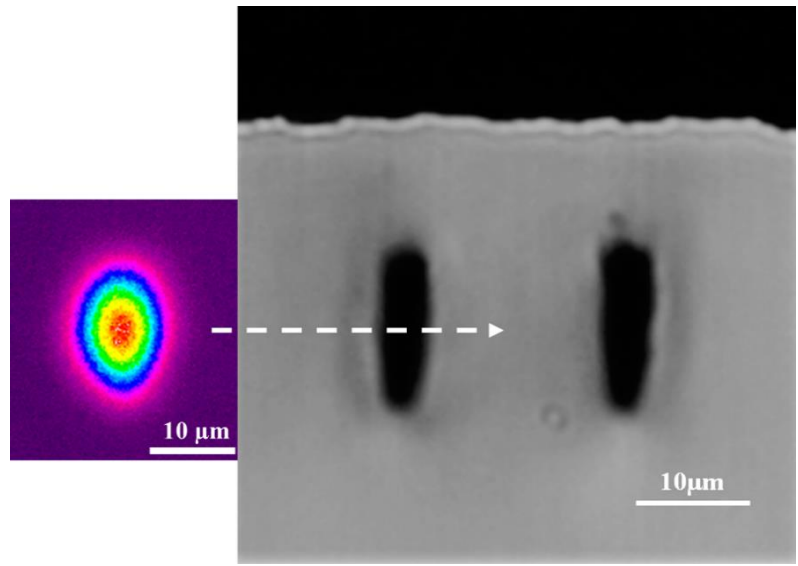

**Figure 1.** Cross sectional microscope image of one facet of the shallow diamond waveguide written at a depth of $d$ = 20 µm (surface to center of modification) and a Type II separation of 19 µm, which exhibits an insertion loss of 12.4 dB. On the left of the microscope image is the near-field mode profile of the waveguide.

An overhead view of the butt-coupling of a single mode fiber (SMF) to a Type II diamond WG is shown in Figure 2a. To further increase the evanescent coupling, it is essential to have a smooth surface finish without any contaminants. The surface roughness is <2 nm, and the diamond microsphere was cleaned with an isopropanol and acetone mixture inside an ultrasonic bath, to remove any surface contaminants. A microscope image of the diamond microsphere, after the cleaning process, is shown in Figure 2b. Moreover, light coupling from an OFHC to the Type-Ib diamond microsphere was previously performed and analyzed [26]. The diamond microsphere resonator is capable of manifesting high Q-factor WGMs; however, a combination of such diamond-based optical components has never been subject to experimentation. In this paper, we report on an all-diamond photonic circuit implementation by integrating a diamond microsphere with a femtosecond-laser-written bulk diamond waveguide for the first time to our knowledge.

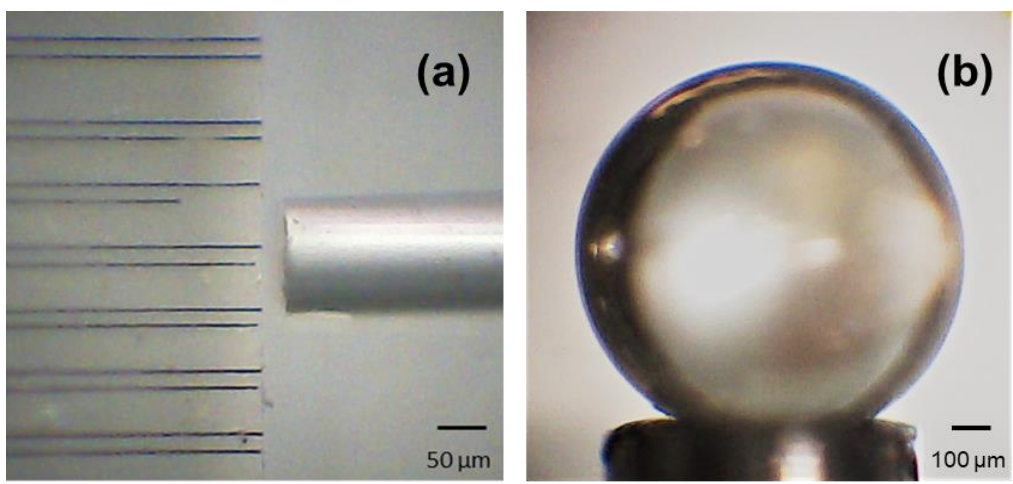

**Figure 2.** (**a**) Single mode bare fiber butt-coupling to a Type II diamond waveguide. (**b**) The 1 mm diameter diamond microsphere being held by a needle tip and a suction mechanism.

Figure 3 illustrates the experimental setup used for observing the elastic scattering of the diamond microsphere on an fs-laser-inscribed diamond WG. The diamond microsphere was placed on the diamond WG by using a needle tip connected to a vacuum pump. As a light source, a narrow linewidth distributed feedback (DFB) semiconductor laser was used at a central wavelength of 1427.7 nm in continuous wave (CW) mode. The DFB laser was adjusted via its thermo-electric control (TEC) control unit to fine tune the laser wavelength with a 1 pm spectral resolution. The DFB laser was further pigtailed to a SMF. The output of the laser was coupled to the WG with a bare SMF, where the polarization was adjusted by a fiber polarization controller. Then, 90° elastically scattered light from the diamond microsphere on the diamond WG was collected with a 10×, 0.42-NA microscope objective to transmit the light to an InGaAs P-I-N photodiode (PD1) attached to a 10× microscope eyepiece through a beamsplitter (BS) and a Glan polarizer (GP). The orientation of the GP with respect to the transmission direction differentiates the TE- and TM-polarized elastically scattered light. Additionally, there is a camera system (Cam) on the other branch of the microscope's collection arm to indicate the position of the diamond microsphere. Moreover, the SMF at the output facet of the WG is connected to a 50/50 optical fiber Y-coupler to monitor the transmitted light. After the fiber Y-coupler, half of the light in the transmission direction was detected by the second photodiode (PD2), and the other half is collected by an InGaAs optical wavemeter (OWM) head controlled by an optical multimeter (OMM), which is further controlled by the control computer (PC). Both PD1 and PD2 are directly connected to a digital storage oscilloscope (DSO) to acquire the 90° elastic scattering and 0° transmission spectra, respectively.

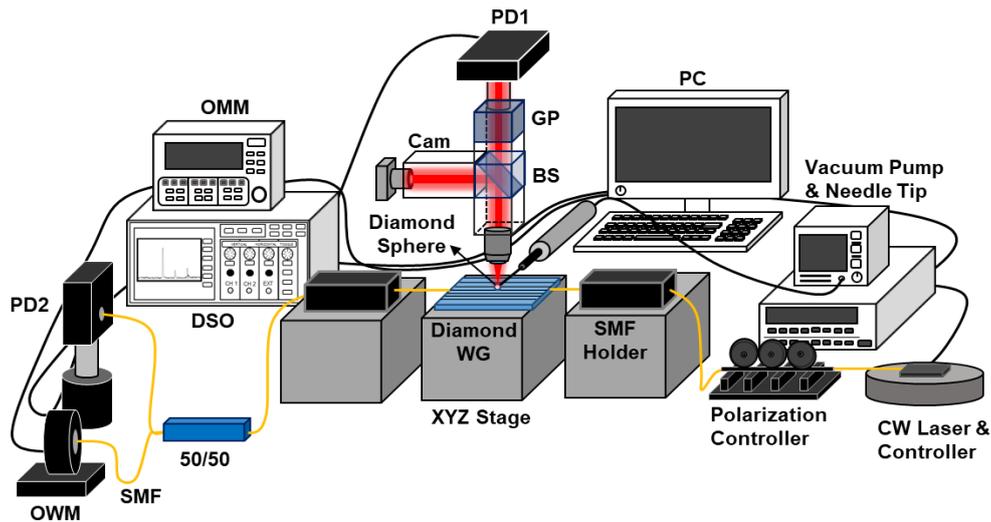

**Figure 3.** The measurement setup to acquire the 90° elastic light scattering and the 0° transmitted optical power to measure the microsphere whispering gallery modes (WGMs) and the waveguide Fabry–Pérot resonances.

In order to realize sufficient coupling of light from the photowritten WG to the diamond microsphere, it is crucial to be within the limits of the impact parameter. One can estimate the impact parameter as $b = 520$ μm, by considering the depth of the WG center from the diamond surface (20 μm) and the diamond microsphere radius (500 μm), which are in contact. The restriction on the impact parameter is formalized as $a \leq b \leq Na$, where $a$ is the radius of the microsphere, and $N = 2.4$, the refractive index of the diamond. Therefore, the maximum calculated impact parameter is $b_{max} = 1.2$ mm, which yields a freedom of less than 680 μm between the surface of the diamond microsphere and the surface of the diamond platform. It is possible to control and manipulate the coupling efficiency and the Q-factor by levitating the diamond microsphere within the defined boundaries.

On the other hand, it is beneficial to characterize the diamond microsphere by estimating the total number of modes. By relating the size parameter to the impact parameter, the number of excited modes can be estimated with $b = (m + 1/2) \times (a/x)$, where $m$ is the mode number, and $x$ the size parameter. According to the localization principle, the size parameter defines the boundaries to the possible number of modes circumnavigating the cavity by $x \leq m \leq Nx$ [27]. The dimensionless size parameter is the ratio of the spherical particle to the wavelength as $x = (2\pi a N_{out})/\lambda$, where $N_{out}$ denotes the refractive index of the surrounding medium, and $\lambda$ the central resonance wavelength at 1427.5 nm. The size parameter is $x = 2200$, and the maximum achievable mode number of spatial WGMs is $m_{max} = 5200$. The estimated number of modes for the corresponding size parameter of 520 μm is $m = 2288$ inside the diamond microsphere.

When WGMs are the subject of interest, it is essential to demonstrate the mode spacing between two consecutive modes of the consecutive mode families. The mode spacing of a resonance inside a spherical cavity is formulated as [28]:

$$\Delta \lambda = \frac{\lambda^2 \tan^{-1}\sqrt{N^2-1}}{2\pi a \sqrt{N^2-1}} \qquad (1)$$

According to Equation 1 the WGM spacing is calculated as $\Delta \lambda = 0.34$ nm for the diamond microsphere. Similarly, the diamond WG will yield a resonance as well, since it can be considered as an FP cavity. The free spectral range (FSR) of an FP resonance [29], similar to the mode spacing, can be defined as $\Delta \lambda_{WG} = \lambda^2/2LN = 86$ pm for the $L = 5$ mm long diamond WG FP resonances with Q-factors on the order of $10^4$.

Figure 4 represent the TE- and TM-polarized 90° elastic scattering and 0° transmission spectra. The WGM spacing of the diamond microsphere was measured as 0.33 nm as indicated in the elastic scattering spectra, which is in close agreement with the estimated mode spacing. Similarly, the FSR of the FP resonance inside the diamond WG was measured as 87 pm and observed in the transmission spectra. Therefore, $\Delta \lambda$ and $\Delta \lambda_{WG}$ measurements verify the WGM and FP resonances inside the diamond microsphere and WG, respectively.

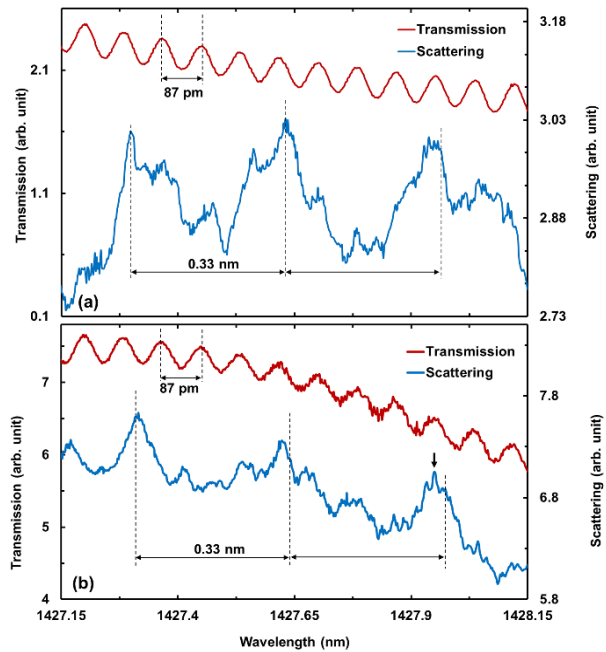

**Figure 4.** (**a**) Transverse electric (TE)- and (**b**) transverse magnetically (TM)-polarized measurement of 90° elastic scattering and 0° transmission spectrum of diamond sphere on diamond waveguide in the near-infrared telecommunication spectral region. The black arrow indicates the highest Q-factor resonance peak.

Furthermore, the most crucial aspect of a resonator is defined by the Q-factor of the resonances. Experimentally, the Q-factor is determined by $Q = \lambda/\delta\lambda$, where $\lambda$ is the resonance wavelength, and $\delta\lambda$ the full-width-at-half-maximum (FWHM) of the resonant Lorentzian peak. The highest measured Q-factor of the FP resonances inside the diamond WG was $10^4$, with an FWHM of 40 pm located around 1427.27 nm. On the other hand, the diamond microsphere yields the highest resonance peak as TM-polarized light with a δλ of 9 pm at 1427.96 nm, as shown in detail in Figure 5 with a Lorentzian curve fitting. This peak is also indicated with a black arrow in Figure 4b. As a result, the maximum measured Q-factor is $1.6 \times 10^5$, comparable with the previous findings of monolithic diamond microsphere WGMs with Q-factor of $2.4 \times 10^7$ [4], and $4.5 \times 10^4$ [26] limited by the respective material losses and the spectral resolution of our measurement setup. As demonstrated in [25], the diamond WG provides better light coupling and propagation in TM mode, thus yielding a higher Q-factor.

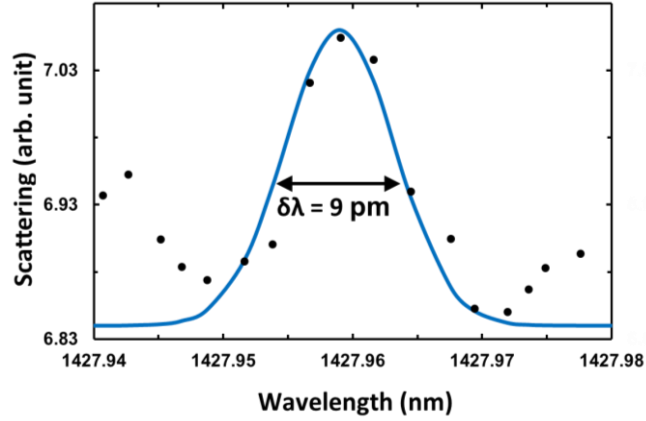

**Figure 5.** The TM-polarized highest Q-factor 90° elastic scattering data (black dots) and Lorentzian curve fitting applied to the data (blue line).

There are several factors affecting the total achievable Q-factor, as shown in:

$$Q_{tot}^{-1} = Q_{rad}^{-1} + Q_{ss}^{-1} + Q_{cont}^{-1} + Q_{mat}^{-1} + Q_{ext}^{-1}, \qquad (2)$$

where $Q_{rad}$ is the loss due to intrinsic radiative loss, $Q_{ss}$ is the scattering loss, $Q_{cont}$ is the loss caused by the surface contaminants, the material loss is $Q_{mat}$, and the loss due to the external light coupling is $Q_{ext}$ [30].

For the diamond microsphere used in our work, $2a/\lambda \gg 15$ means that $Q_{rad}$ can be omitted [31]. Likewise, based on the manufacturing process for the diamond microsphere, $Q_{cont}$ is negligible [26]. The scattering losses are due to the surface inhomogeneity and estimated by $Q_{ss} = (\lambda^2 a)/(\pi^2 \sigma^2 B)$ [30]. Here, $\sigma$ is the rms size, and $B$, the correlation length of the surface inhomogeneity, both of which are estimated as <2 nm by taking the applied cleaning process into account [30]. As a result, $Q_{ss}$ is found to be $10^{10}$, thus not limiting the Q-factor.

The components, that have the most significant impact on the Q-factor of the WGMs in the diamond microsphere, are $Q_{mat}$ and $Q_{ext}$. The material loss is defined as $Q_{mat} = (2\pi N)/(\alpha \lambda)$, where $\alpha$ is the attenuation coefficient, in the order of 1 cm$^{-1}$ [32]. $Q_{mat}$ is calculated to be in the order of $10^5$.

In addition, the loss caused by the external light coupling can be estimated by $Q_{ext} = 2\pi x/|t^2|$ [33], where $t$ is the incident light mode field-coupling coefficient from the fiber to the diamond microsphere. $|t^2|$ is the energy loss per round trip and depends on the spatial beam form propagating inside the diamond WG. $|t^2|$ is correlated with the coupling efficiency, which can be estimated with $(\sqrt{2}\lambda)/(\pi 2\omega_o)$, where $\omega_o$ is the $1/e^2$ width of the MFD of the beam inside the diamond WG [34]. Based on our experimental parameters, the coupling efficiency is estimated as ~3.3% [25]. As a result, $Q_{ext}$ is calculated to be in the order of $10^5$. It is possible to increase the coupling efficiency by carefully tailoring the beam profile during the manufacturing of the diamond WG.

In light of these findings, $Q_{mat}$ and $Q_{ext}$ are the factors that determine the limits to the possible achievable Q-factor. It is possible to increase the Q-factor by selecting a diamond microsphere with lower absorption for this selected wavelength regime, which is relatively less troublesome to fabricate nowadays with the developing CVD manufacturing technology [32]. On the other hand, by minimizing the external coupling losses with the optimization of the beam profile inside the diamond WG, it will be possible to achieve Q-factors in the order of $10^6$.

## 3. Materials and Methods

The diamond WGs were written by a Yb:KGW femtosecond (fs) pulsed laser (Pharos, Light Conversion, Vilnius, Lithuania) with a 515 nm central wavelength, 500 kHz repetition rate, and 230 fs pulse duration. The ultrashort pulses were focused by a 100× oil immersion objective with a 1.25 numerical aperture (NA) below the surface of the diamond. The WGs were photo-inscribed inside an optical grade diamond (MB Optics) sample with nitrogen impurities of ~100 ppb, having dimensions of 5 mm × 5 mm × 0.5 mm. The laser power was varied between 30 mW and 40 mW, with 19 μm spacing between the tracks forming the Type II WG. The waveguides were written with the same scan speed of 0.5 mm/s, with 50 μm spacing between successive Type II WGs.

The Type-Ib (nitrogen impurity >5 ppm) diamond microsphere (Dutch Diamond Technologies, Cuijk, The Netherlands) has a 1 mm diameter. The microsphere was grown in a laboratory environment via chemical vapor deposition (CVD). The form accuracy defining the roundness of the microsphere is <250 nm, which is achieved by lapping.

## 4. Conclusions

In this work, the WGM excitation with both TE- and TM-polarized light of a 500 μm radius diamond microsphere coupled to an fs-laser-written diamond WG was demonstrated. We realized microsphere WGM resonances with Q-factors of $10^5$. It is possible to achieve high-resolution resonant peaks by using high-quality CVD diamond with minimal absorption losses and by optimizing the MFD inside the diamond WG. Spherical diamond microresonators on fs-laser-written diamond WGs show promise as novel all-diamond integrated photonic architectural components. By carefully engineering these high-quality factor resonances and further exploiting the properties of existing nitrogen-vacancy centers in diamond microspheres and diamond waveguides in such configurations, it should be possible to realize filtering, sensing and nonlinear optical applications in integrated diamond photonics.

**Author Contributions:** Conceptualization, A.S., R.R. and S.M.E..; methodology, N.Y., M.M.B., H.O.Ç., T.L.P., A.G., V.B.; writing—original draft preparation, N.Y.; writing—review and editing, A.S.; S.M.E.; funding acquisition, S.M.E., A.S. All authors have read and agreed to the published version of the manuscript.

**Funding:** This work was supported by Türkiye Bilimsel ve Teknolojik Araştırma Kurumu (TUBITAK) [grant number 114F312] and CONCERT-Japan DiamondFab Project, European Commission (EC) [grant number FP7-INCO-266604]. The research contracts for H. O. Çirkinoğlu and M. M. Bayer were supported by Türkiye Bilimsel ve Teknolojik Araştırma Kurumu (TUBITAK) Grant No: 114F312. S. M. Eaton was supported by the DIAMANTE MIUR-SIR grant and FemtoDiamante Cariplo ERC reinforcement grant. T. Le Phu and A. Giakoumaki were funded by the H2020 Marie Curie ITN project PHOTOTRAIN. A. Giakoumaki is grateful for support from the research project sPATIALS3 (Regione Lombardia). V. Bharadwaj is grateful for financial support from the ERC project PAIDEIA GA n.816313.

**Acknowledgements:** We thank the Dr. Luigino Criante and Prof. Guglielmo Lanzani Center for Nano Science and Technology-Italian Institute of Technology (CNST-IIT) in Milano for access to the FemtoFab laser fabrication facility.

**Conflicts of Interest:** The authors declare no conflict of interest.

9 of 9